# Nanoscale devices with superconducting electrodes to locally channel current in 3D Weyl semimetals

Biswajit Datta,[1, a)] Jaykumar Vaidya,[1, a)] Subhamoy Ghatak,[1] Raghav Dhingra,[1] Rajib Mondal,[1] John Jesudasan,[1] A. Thamizhavel,[1] and Mandar M. Deshmukh[1, b)]

*Department of Condensed Matter Physics and Materials Science, Tata Institute of Fundamental Research, Homi Bhabha Road, Mumbai 400005, India*

We report on the fabrication of nano-devices on the $[\bar{1}\,0\,1]$ surface of a Weyl semimetal, a macroscopic crystal of TaAs, and low-temperature transport measurements. We can implement electron beam lithography by peeling off and transferring the resist for nanofabrication onto the irregular crystal. We fabricate the device electrodes with superconducting Niobium nitride (NbN) to control the current flow through the intended active area of the devices. Our device structure enables the reduction of the current jetting effect, and we demonstrate the negative magnetoresistance measurement as a function of angle. The high field magnetotransport show three distinct oscillation frequencies corresponding to the three bands at the Fermi level. Resistance measured in the low magnetic field shows the usual weak anti-localization dip near the zero-field – a signature of a Weyl material. Our method of fabricating devices with superconducting electrodes provides a way to probe the electrical properties of macroscopic single crystals at the nanoscale. As we use conventional lithographic techniques for patterning, this method can be extended to a wide gamut of electrode materials and a large class of 3D quantum materials.

a)equal contribution

b)deshmukh@tifr.res.in

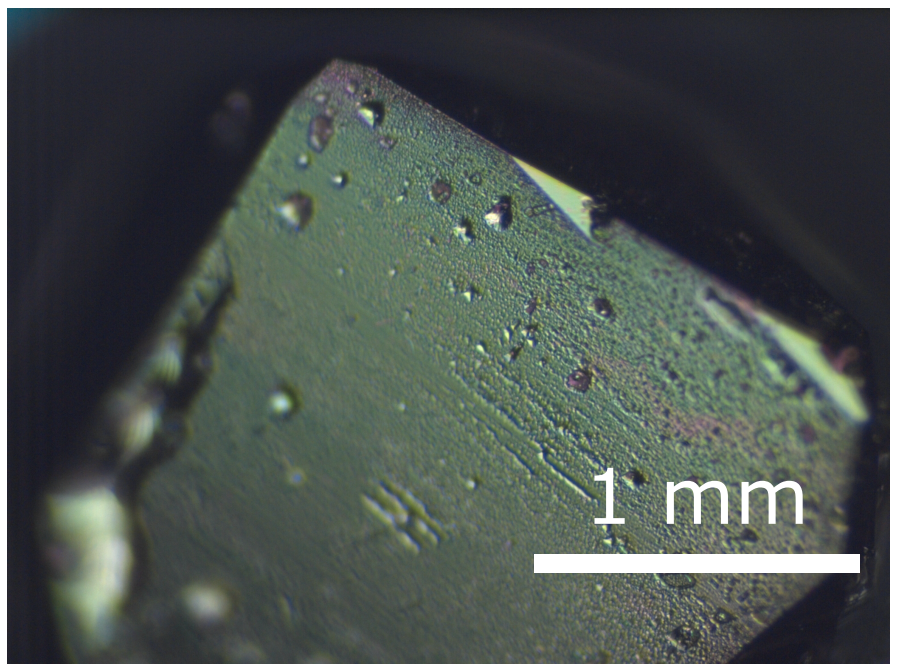

FIG. 1. **Optical image of the crystal**. The topography shows that the crystal surface is not smooth everywhere – we carefully choose areas to make devices avoiding visible defects.

There is an increasing interest in quantum materials with unique topological properties. In order to probe the properties, one needs to study high-quality single crystals. Various growth techniques are used to make the single crystals, and they often lead to irregularly shaped crystals and irregular surfaces. Unlike 2D and other conventional materials, like Si wafer, close to atomically flat, lithographic techniques for making nano-devices on single crystals are hard to implement.

Weyl semimetals[1] provide an exciting platform to study massless Weyl electrons. Besides hosting topological conductive surface states connecting the projection of the bulk Weyl nodes, the bulk band structure of Weyl semimetals also carries unique signatures[2,3]. One such signature is chiral anomaly[4,5] – the nodal charge imbalance in the presence of a parallel electric and magnetic field for which the primary experimental signature is the negative magnetoresistance. Another signature of chiral anomaly is detecting the non-local voltage induced by the nodal (valley) current[6]. This requires making tiny devices as valley relaxation length is only a few micrometers[6].

This report demonstrates a method that enables making nanoscale precision devices on irregular tiny bulk single crystals where conventional lithography does not work. Typically, attaching leads with silver epoxy paste is a popular method to attach electrical contacts on bulk crystals for measuring electrical properties. While it is simple, it is sometimes hard to make quantitative measurements, and the precision is limited due to the manual alignments of the electrical leads. On the other hand, electron beam lithography makes high precision devices, but it is hard to implement on bulk crystals due to uneven surface and their typical small sizes (few $mm^2$ for TaAs). In addition, as we discuss later, "current jetting", in the presence of a magnetic field, poses challenges in high-mobility crystals. We do electron beam lithography with peeling off resist to solve this problem, which allows us making high precision devices. While some techniques based on resist peeling have been implemented[7–9], our technique allows more flexibility. We also demonstrate the use of resist transfer technique to make devices with single crystal quantum materials. This technique crucially relies on peeling and subsequent resist transfer from a flat Si substrate and following it with conventional e-beam based lithographic fabrication. Focused ion-beam[10] has also been used by some groups to fabricate mesoscopic devices of bulk crystals of quantum semimetals, but that method can introduce defects due to high energy ions used in FIB. As we do not use a focused ion beam (FIB) to cut the crystal, no damage to the single crystal occurs. We can deposit any material that can be deposited by evaporation and sputtering.

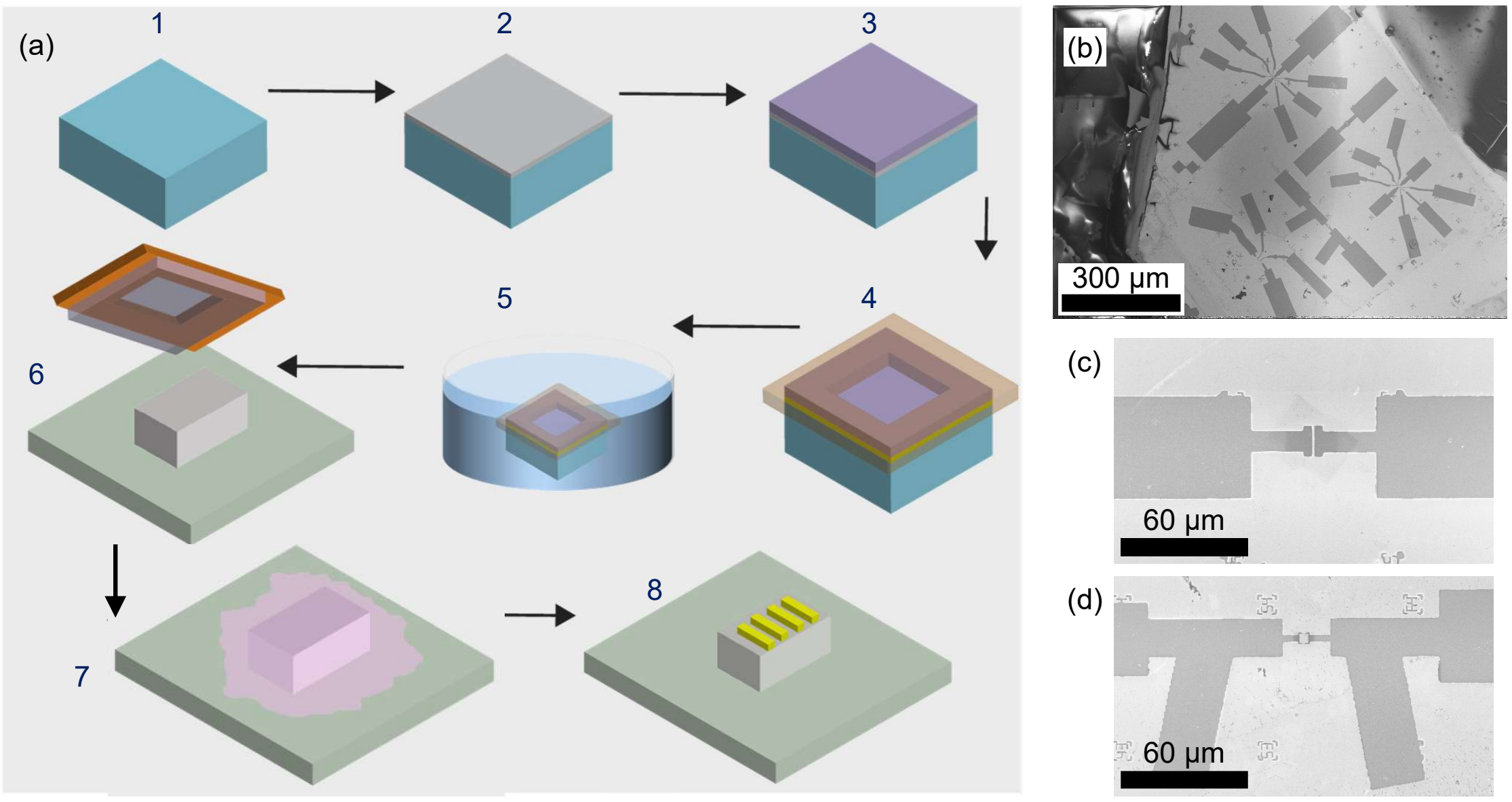

FIG. 2. **Device and fabrication process.** (a) 1, A bare silicon chip that is used to spin uniform resist in the next step. 2, Spin coating a blank Si/$SiO_2$ chip with the PVA and then desired e-beam resist as shown in step 3. 4, Sticking the punched Kapton tape on the chip, immersing the chip into the DI water as shown in step 5. The Kapton tape holding the resist gets separated from the chip and floats upon the water. 6, Removing the Kapton tape holding the resist from the DI water. 7, Gently placing the resist on the crystal of interest followed by soft baking at 180 C for a minute. 8, patterning metal electrodes using e-beam lithography and deposition of metal. (b) SEM image of the completed devices fabricated on the surface of the crystal with the method described before. (c) and (d) show the short channel with wide electrodes geometry and two-prong geometry, respectively.

Fig. 1 shows the optical image of a typical TaAs crystal surface (details of crystal growth and characterization provided in section 1 of Supplementary Material). The imperfections are visible even in an optical image. It is important to avoid imperfections to make continuous electrodes. The bulk band structure[2,3] of TaAs has 12 pairs of Weyl points in which 4 pairs are at $k_z = 0$ plane – termed as $W_1$ Weyl points. Remaining 8 pairs are offset from $k_z = 0$ – termed as $W_2$ Weyl points. Among all $W_1$ and $W_2$ Weyl points, only one pair of $W_1$ and one pair of $W_2$ Weyl points are independent; all other Weyl points are related with the reciprocal lattice vectors. In addition to the non-trivial linear bands, there is also a trivial hole band at the Fermi energy.

In order to fabricate spatially resolved very short channel length ($< 1\mu$m) devices, we need to pattern electrodes using a high-resolution patterning process like electron beam lithography. TaAs bulk crystals are rough and irregular in shape – hence it is difficult to pattern the bulk crystal using the conventional lithography method, which involves spin-coating the electron beam resist, and the resist thickness can vary significantly near rough areas and edges. For uniformity of resist, which dictates minimum feature size, a flat surface is required. The essence of our technique is to use water-soluble polyvinyl alcohol (PVA) as a sacrificial layer coated before multiple layers of e-beam resist on a flat Si substrate. We later dissolve the PVA to peel layers of dry resist, which is then used to coat the crystal. The uniformity of the resist on the crystal is as good as on a flat substrate.

In the following, we briefly describe the process. We glue the crystal on a Si substrate for ease of handling with a strong thermally conducting epoxy (stycast) that can mechanically hold the crystal throughout the fabrication process. Before this, the orientation of the crystal face is determined using Laue diffraction to identify the surface of interest. A thin layer of PVA solution (Poly Vinyl Alcohol PVA: water = 1g:19g) is spin-coated on a Si chip with dimensions $\sim 1$ cm $\times$ 1 cm or larger. The size of the chip may change depending on the area of the resist one intends to peel. Spin coating of the desired e-beam resists – in this case, two layers of PMMA 950 A4 (total thickness  400 nm) are done on top of the PVA layer on the same Si chip. Using a biopsy punch, we make a hole in a Kapton tape. This serves as the frame for picking up resist. We stick the Kapton tape on the chip containing the PVA/resist multilayer and immerse that in the preheated DI water at 120 C for about 5 minutes. We can then easily peel off the Kapton tape as it detaches from the chip after the PVA layer gets dissolved in DI water. The e-beam resist can be visually seen clamped on the Kapton tape frame after removing it from the DI water. We stick the Kapton tape containing the clamped resist on the desired area of the TaAs crystal. Necessary alignments can be made such that the resist is transferred over the uniform area. The TaAs crystal is baked at 180C for 2 min so that the resist adheres to the crystal surface uniformly. E-beam lithography is done on the crystal to pattern desired device structures. After e-beam lithography and development, NbN is deposited using sputtering. In the end, a liftoff process is used to get the electrodes for nano-devices.

Fig. 2a shows the schematic of the various steps in the fabrication process. Fig. 2b shows a low magnification SEM image of all the devices fabricated on a TaAs surface. Fig. 2c and Fig. 2d show the SEM image of a wide electrode short-junction and two-prong device geometry respectively.

The conducting surface of Weyl semimetals (or any other topological metal), like TaAs, poses another challenge in defining isolated devices on the surface. The whole surface is highly conductive, making it hard to channel the current through the intended area if standard electrodes like Ti/Au are used to make electrodes on TaAs. To overcome this problem, we use the superconducting NbN to make the electrodes. The high critical field (>16 T) and $T_C = 16$ K ensures that the electrodes remain superconducting even at the highest field of our measurement (14 T). Our choice of electrode material is vital as most of the current flow via NbN till the edge of the patterned electrodes and then flows through Weyl semimetal for the shortest possible path, reducing the shunting of the device with highly conducting bulk crystal.

Since we deposit a thin film of NbN to define the electrodes for nano-devices, it is essential to ensure that the film retains its superconductivity. To check whether there is any inverse-proximity[11] on the NbN film due to TaAs, we deposit a 580 nm wide and 150 nm thick NbN wire on the TaAs crystal and measure its resistance at 1.5 K in four probe geometry. Fig. 3a and Fig. 3b show an SEM image of the NbN wire and its resistance as a function of bias DC and field. We see that even for 580 nm wide electrodes and 150 nm thick film, the critical current at 300 mK remains $\sim 10$ μA at 14 T magnetic field. The electrode width in our TaAs devices is few μm which keeps the leads superconducting with a much larger critical current. Confirming that the inverse proximity effect is not taking place ensures that the injection of current is local in our devices. Our results are consistent with previous studies of inverse proximity effect on NbN[12].

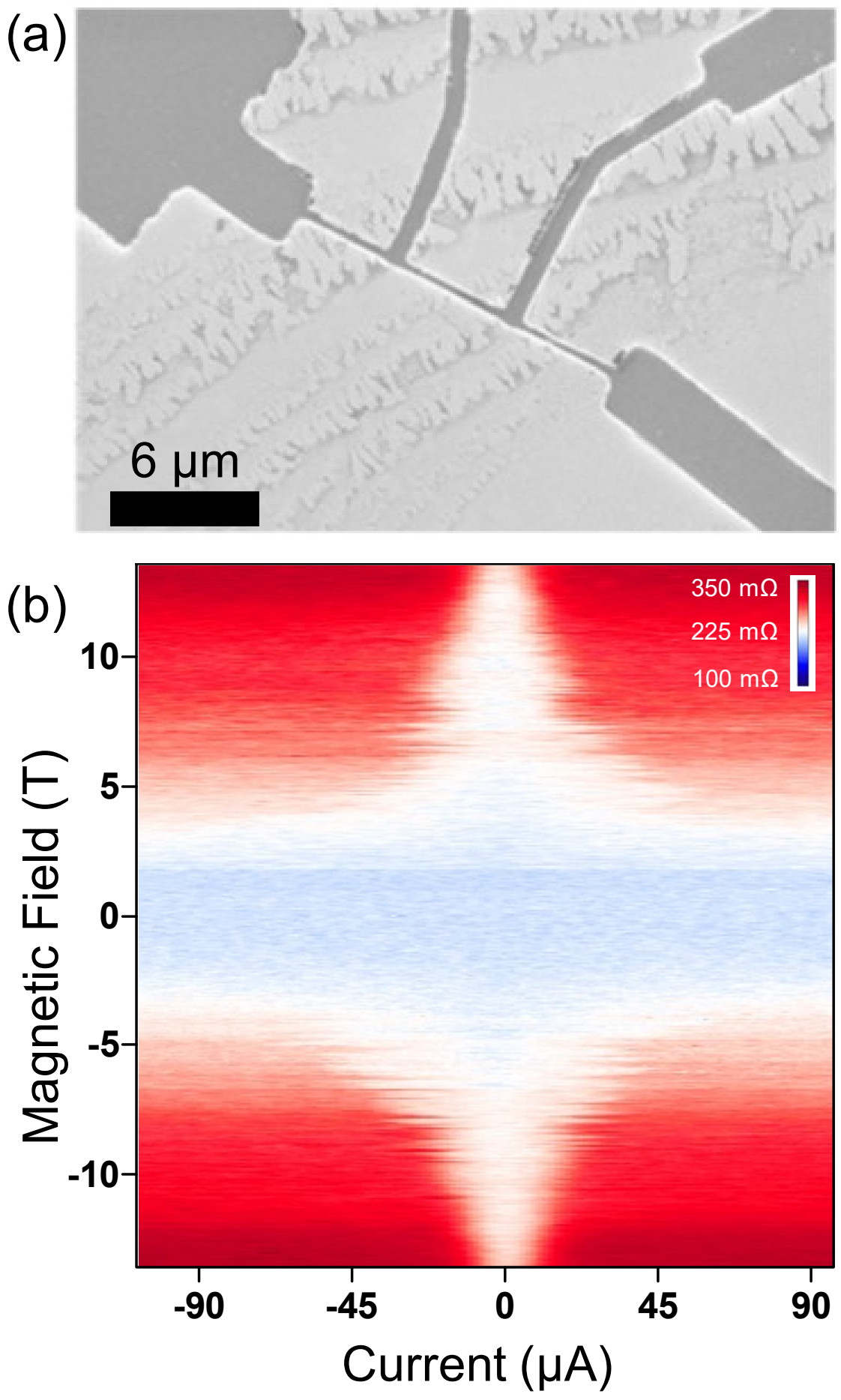

FIG. 3. **Test for the inverse proximity effect in the thin NbN superconducting film on TaAs.** (a) SEM image of a 580 nm wide and 150 nm thick NbN wire deposited on TaAs to check whether it retains superconductivity. (b) Color scale plot of the differential resistance of the NbN wire as a function of bias DC and magnetic field at 300 mK. The blue color shows the area of the plot where the NbN wire is superconducting.

Magnetotransport measurements are important to probe the band structure and Fermi surface properties[13,14]. One of the key aspects of magnetotransport measurements in high mobility bulk crystals is that due to a magnetic field, the current density is non-uniform and can result in anomalous voltage drop. This aspect is also known as current jetting[15–20], and occurs when locally the current flow direction is not parallel to the voltage electrodes. The magnetoresistance voltages measured by two probe sets on mutually perpendicular faces of the specimen are observed to have a significant difference. Our technique of fabricating superconducting electrodes to inject current locally on the surface and fabricating voltage probes on the same surface offers a quasi-2D geometry that suppresses current jetting (details of finite element simulations provided in section 2 of Supplementary Material). Here, the separation between electrodes on the surface is much smaller than the dimensions of the crystal.

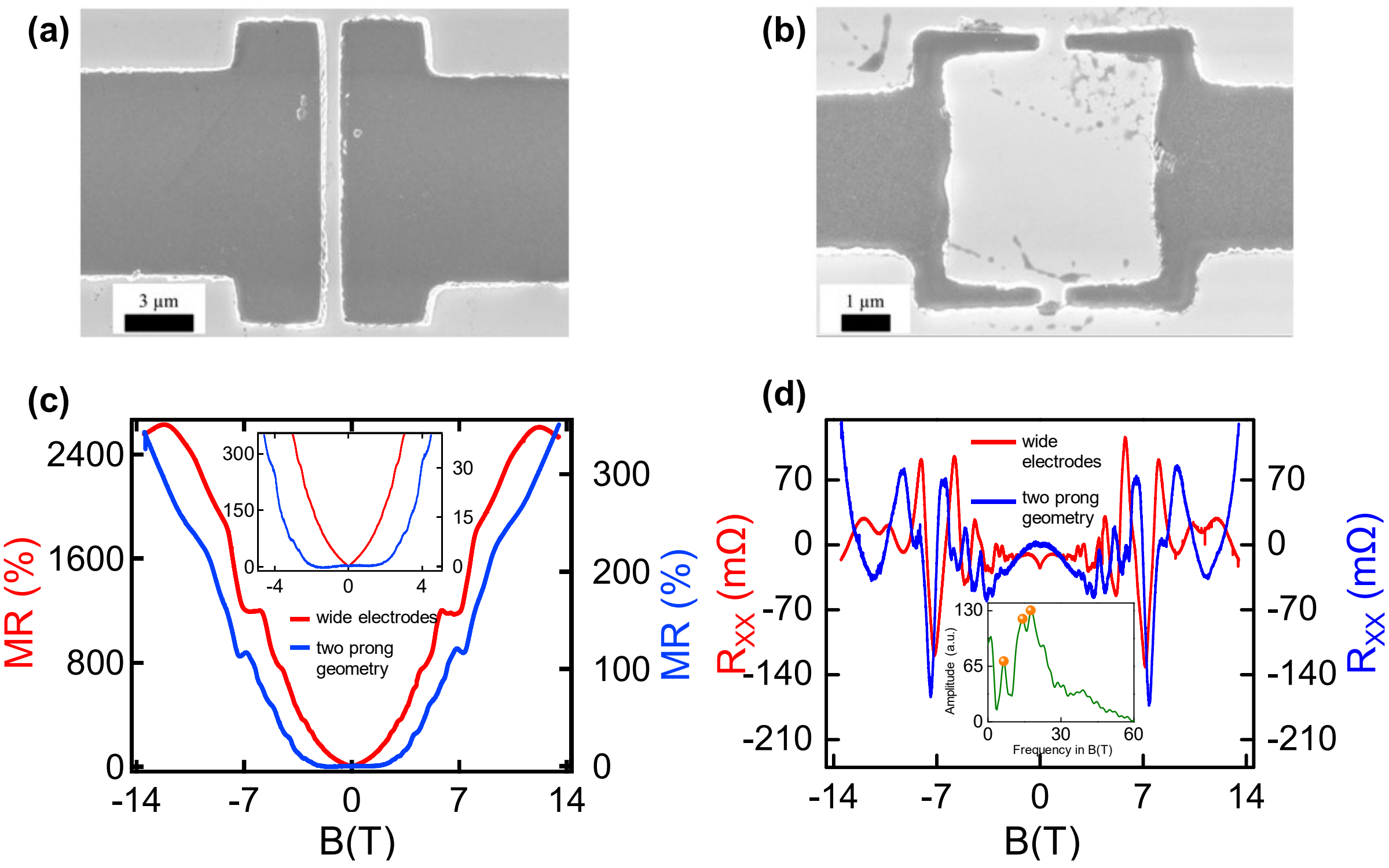

FIG. 4. **Comparison of the short channel and two prong geometry.** (a) and (b) show the zoomed-in SEM image of the devices in wide electrodes short channel geometry and two prong geometry respectively. (c) Longitudinal magnetoresistance of the two devices measured at 1.5K. Inset shows Zoomed-in magnetoresistance at low field. The magnetoresistance for the two-prong geometry in $-3$ T $< B <$ 3 T changes non-monotonically. On the other hand, wide electrodes short channel geometry shows usual weak localization WAL at low field. (d) After background subtraction from the magnetoresistance data shown in panel-c measured at 1.5K. Inset shows the Fourier transform of the background subtracted data of the wide electrodes short channel geometry device showing three main FFT peaks – 6.6 T, 14.4 T and 17.7 T. These three SdH frequencies are attributed to two Weyl nodes $W_1$, $W_2$ and the trivial hole band.

We made two different device geometries – Fig. 4a shows a short channel with a wide electrode device with $\sim$ 590 nm long TaAs channel and Fig. 4b shows a device with "two prong" geometry. For both of these geometries, the current and voltage electrodes are the same – this ensures that the voltage electrodes are parallel to the flow of current, and hence the current jetting effect is minimized. In both the devices, we wirebond directly onto the superconducting bonding pads to inject the current into the crystal only at the end of the electrode tip. The current distribution is expected to be more uniform for the wide electrode geometry as the current is injected uniformly in the entire channel. We show the magnetoresistance of these two devices in Fig. 4c. Fig. 4c inset shows the zoomed-in part of the data at low field. The short channel device shows a cusp ("V" shape) in the resistance near zero field – a signature of the weak anti-localization in a Weyl semimetal. On the contrary, the two-prong geometry at low field interestingly shows non-monotonic magnetoresistance. Fig. 4d shows the Shubnikov-de Haas (SdH) oscillations after subtracting a smooth background. The contrast between WAL (for the wide electrode geometry) and non-monotonic magnetoresistance (for the two-prong geometry) is better visible in the background-subtracted data. Between these two geometries, there exists almost a $\pi$ phase difference which is also apparent at the high field SdH oscillations – oscillation peaks of the short channel device roughly coincide with the valleys of the double prong geometry, seen in Fig. 4d. We note that the WAL in wide electrode geometry manifests the expected $\pi$ Berry's phase of the Dirac bands in TaAs. This means the two-prong geometry picks up another $\pi$ phase whose origin is not clear to us. However, it turns out that the geometry of electrodes can contribute to the phase difference. It has been observed that the WL and WAL features can be modified by the geometry of the device[21]. A part of the change in response could result from the physics due to the competition of length scales. The ability to make controlled geometry of electrodes, as we show, in topological material could enable the exploration of new physics.

Fig. 4d inset shows the Fourier transform of a smooth background-subtracted magnetoresistance of the short channel wide electrode device. In this crystal, the Fermi level is such that we can see all the three SdH frequencies. The FFT frequencies that we get from our data are 6.6 T, 14.4 T, and 17.7 T. These values are similar to the values (7 T and 19 T for B || [0 0 1] direction) reported before[22] and correspond to the Weyl nodes $W_1$, $W_2$, and a trivial hole band[22]. Similar frequencies have also been reported[23] for bulk Fermi pockets (7.3 T and 19.9 T for B || [0 0 1] direction). However, the exact values are not expected to match as we have fabricated the devices on [-1 0 1] surface which is different from their studies.

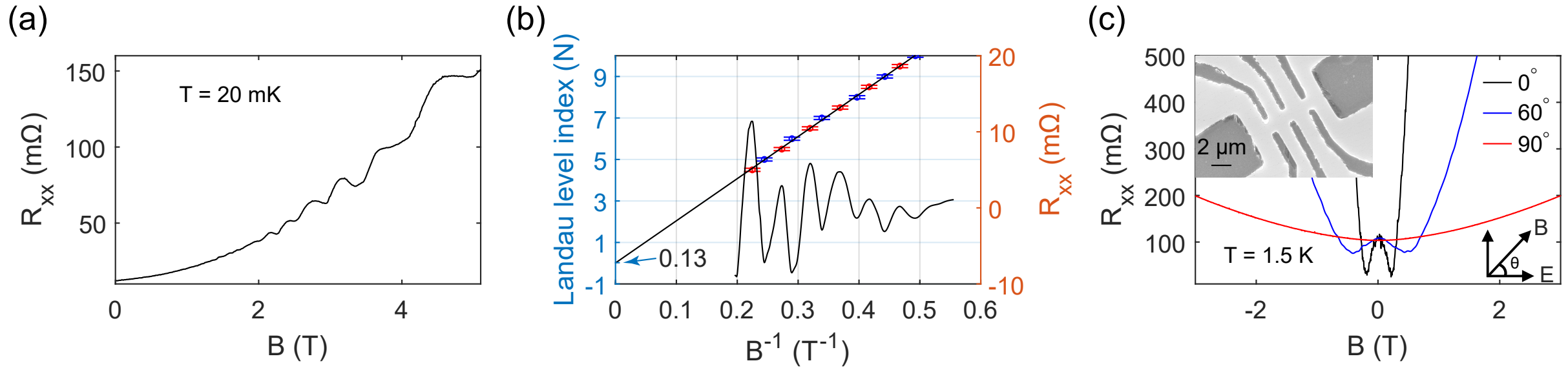

FIG. 5. **Phase of the oscillations and negative magnetoresistance.** (a) Longitudinal magnetoresistance from a device where primarily one frequency is resolved. (b) The blue curve is the background-subtracted longitudinal magnetoresistance plotted with the inverse magnetic field. Integer (half-integer) LL indices are assigned to the maxima (minima) of the oscillations to produce the LL index plot. The intercept is $0.13 \pm 0.07$, which exhibits $\pi$ Berry's phase of the Weyl node. (c) Longitudinal magnetoresistance for different values of the angle between electric field (direction of current flow) and magnetic field. Inset shows the SEM image of the measured device.

We further study the SdH oscillations to reveal Berry's phase of the non-trivial bands. Lifshitz-Onsager quantization rule for any closed orbit in k-space with area $(A_N)$ can be written as $A_N \frac{\hbar}{eB} = 2\pi(N + \frac{1}{2} - \frac{\Phi_B}{2\pi} + \delta)$ where $\hbar$, $e$, $B$, $N$, $\Phi_B$ are reduced Planck's constant, electronic charge, magnetic field, Landau level index and Berry's phase respectively; $\delta = \pm 1/8$ is an additional phase factor in 3D. Fig. 5a shows the raw data of the SdH oscillations for a device where we primarily resolve a single frequency in the plotted magnetic field range. Fig. 5b shows the background subtracted SdH oscillations and the superimposed LL index plot. We assign the integer LL indices in the oscillation valleys and half-integer indices in the oscillation peaks. A straight line fitted to such a plot produces $\delta = \pm 1/8$ intercept in the $B^{-1}$ axis for non-trivial $\pi$ Berry's phase. Our plot produces an intercept $0.13 \pm 0.07$, which is consistent with the expected $\pi$ Berry's phase in 3D. We also measured magnetoresistance in the presence of the tilted magnetic field. Fig. 5c shows the longitudinal magnetoresistance for different tilting angles. We see that the magnetoresistance is close to a parabola for $E \perp B$, but as we tilt the magnetic field towards the electric field (to make $E \parallel B$), a negative magnetoresistance appears. This is a sign of chiral anomaly[4,5] measured by electrical transport.

To summarize, in this study we demonstrate a reliable fabrication process to make nanoscale devices on single crystals of quantum material and probe its electrical properties. Making nanoscale devices on bulk crystals is essential to do more quantitative measurements and to reduce the current jetting effect in high mobility crystals as we show by finite element simulations (discussed in section 2 of the Supplementary Material). The finite element simulations bear out the fact that our method of making devices on the surface of the single crystals is appropriate for such 3D crystals. Our devices with wide current injecting electrodes, nanometer-scale probes, and superconducting electrodes with the high critical magnetic field ($H_c$) are appropriate for a wide variety of studies of quantum crystals including Weyl semi-metals, as we demonstrate in the manuscript. It is clear that using the described technique, we can make nanoscale electrodes to single crystals of quantum materials using a wide gamut of materials that are available for evaporation and sputtering processes. In addition, ferromagnetic electrodes can be used to probe properties of ferromagnetic Weyl systems[24–26]. It will also be exciting to make lithographic structures like Josephson junctions and SQUIDs on Weyl systems[27] using our method.

## DATA AVAILABILITY

The data that support the findings of this study are available from the corresponding author upon reasonable request.

## ACKNOWLEDGEMENTS:

We thank Ulhas Vaidya and Pratap Raychaudhuri for their experimental assistance. The authors acknowledge the Department of Atomic Energy of the Government of India grant 12-R & D-TFR-5.10-0100, DST Nanomission grant SR/NM/NS-45/2016 and DST SUPRA SPR/2019/001247 grant.

## CONFLICT OF INTEREST

The authors have no conflicts to disclose.

**Supplementary Material**

# Nanoscale devices with superconducting electrodes to locally channel current in 3D Weyl semimetals

## Section 1

## Details about TaAs crystal growth and characterization

The single crystals of TaAs were grown by chemical vapour transport method in a horizontal three-zone furnace. The precursors, powders of Tantalum (purity 99.98 %) and Arsenic pieces (purity 99.9999 %) in the stoichiometric ratio, along with granular pieces of Iodine were vacuum sealed in a quartz ampoule of length of about 250 mm and inner diameter of about 18 mm. Here, iodine acts as the transporting agent for crystal growth. Typically, about 3–4 mg/cc iodine yields a reasonably good size single crystal. The quartz tube with the constituent elements was sealed in a vacuum of $10^{-6}$ mBar. The sealed ampoule was then placed in a horizontal zone furnace, and all the zone temperature was gradually raised to 1050°C over the period of about three days at a rate of heating 15°C/h. All the zones were kept at 1050°C for about 24 h for the formation of a homogenous mixture of TaAs. After homogenization, one of the zones was reduced to 950 °C, thus maintaining a temperature gradient of 100 °C to enable the transport. This temperature gradient in the zone furnace was maintained for 10 days, after which the furnace was cooled down to room temperature. Shiny crystals of TaAs with varied dimensions (~ 1mm × 1mm) were obtained. The crystal composition was obtained using energy dispersive x-ray analysis, which was found to be in the ratio 1:1 of Ta:As. The Single crystalline nature of grown TaAs was checked using Laue diffraction patterns.

## Section 2

## Details about finite element simulation to support the choice of geometry of devices

The finite element simulations bear out the fact that our method of making devices on the surface of the single crystals is appropriate for such 3D crystals. Our devices with wide current injecting electrodes, nanometer scale probes, and superconducting electrodes with high $H_c$ are appropriate for wide variety of studies of quantum crystals including Weyl semimetals, as we demonstrate in the manuscript.

Our simulations (which does not model the chiral anomaly) prove that our devices in the present geometry do not exhibit angle-dependent non-monotonic magnetoresistance. Thus this geometry is appropriate for probing physics of Chiral anomaly. We include anisotropic transport properties and high quantum mobility (from experimental data) in our simulations. As a comparison, we also simulate the devices in the “conventional” geometry and find that the conventional geometry indeed shows strongly angle-dependent non-monotonic magnetoresistance. Thus the conventional geometry can complicate the observation of the Chiral anomaly while our device geometry does not.

We set up various 3D models to self-consistently calculate voltage and current distributions using the finite element software package Comsol Multiphysics 5.5. The simulations are based on solving the stationary Laplace equation and Ohm’s law for current conservation throughout our geometry. The 3D models consisted of superconducting electrodes placed on top of a metallic cube of side 1 mm. The anisotropic resistivity was computed using the Drude model [Magnetoresistance in Metals by A.B. Pippard], with the zero-field resistivity along the a axis taken to be 2.86 x $10^{-8}$ Ohm・m and the electron mobility as 0.5 $m^2(V・s)^{-1}$. The anisotropic resistivity along the c axis was taken to be three times larger than that along a axis, as in the case of TaAs (Xiang *et al* 2017 *J. Phys.: Condens. Matter* **29** 485501).

Fig. S1 below illustrates two different device geometries and their corresponding magnetoresistance curves. The voltage was measured at the points in the middle of the voltage electrodes. We've also tried measuring the voltage by averaging the voltage values over the contact area of the voltage electrodes, and we found the results to be sufficiently similar.

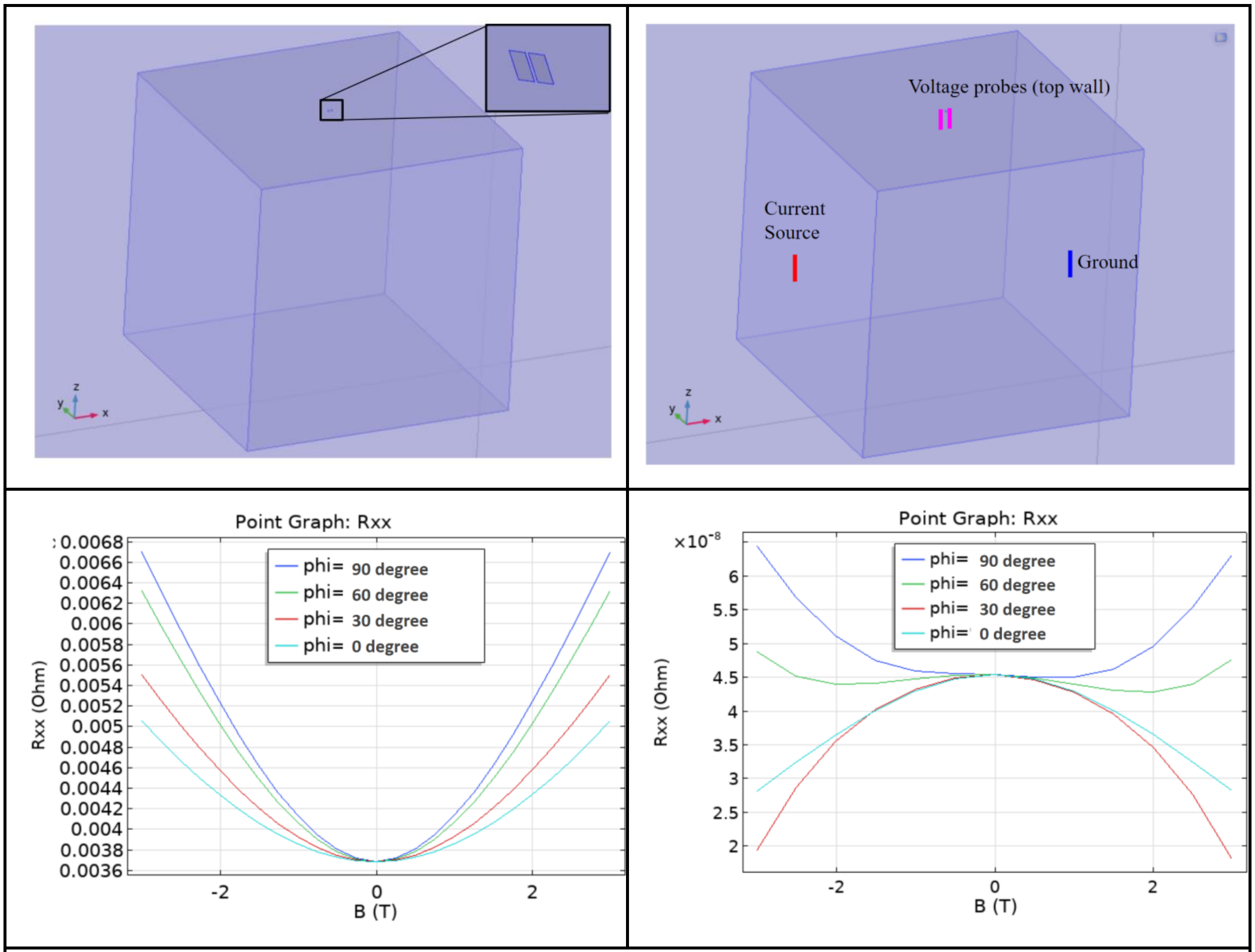

*Figure S1: a) A 3D model showing a pair of electrodes on top of a metallic cube with anisotropic magnetoresistance. The geometric parameters are similar to the device used in our experimental study. b) A 3d model with the current source and ground electrodes placed on opposite faces of the metallic cube. While the voltage probes remain on the top face as in a). This geometry mimics conventional*

*devices on bulk crystal. c) The longitudinal magnetoresistance for the 2-probe device geometry shown in a). 'Phi' denotes the angle between the magnetic field and the current flow. d) The magnetoresistance as measured at the voltage probes for the 4-probe device geometry shown in b).*

The results clearly indicate a positive and significantly more uniform magnetoresistance for shorter channel length as compared to the conventional electrode configuration used for such devices (Fig. S1b). The relative change in Rxx is also small enough in the shorter channel geometry for detecting the signature of the chiral anomaly in Weyl semimetals i.e. negative magnetoresistance, if it were present. Furthermore, the magnetoresistance is non-monotonic for the conventional electrode configuration (Fig. S1d) as a result of current jetting. This reinforces our claim that the shorter-channel geometry is less plagued by current jetting and superior to the conventional geometry.

We also did a simulation to compute the magnetoresistance for a geometry similar to the one shown in Fig. S5c of our manuscript. The trends were found to be consistent with the data shown in Figure S1.

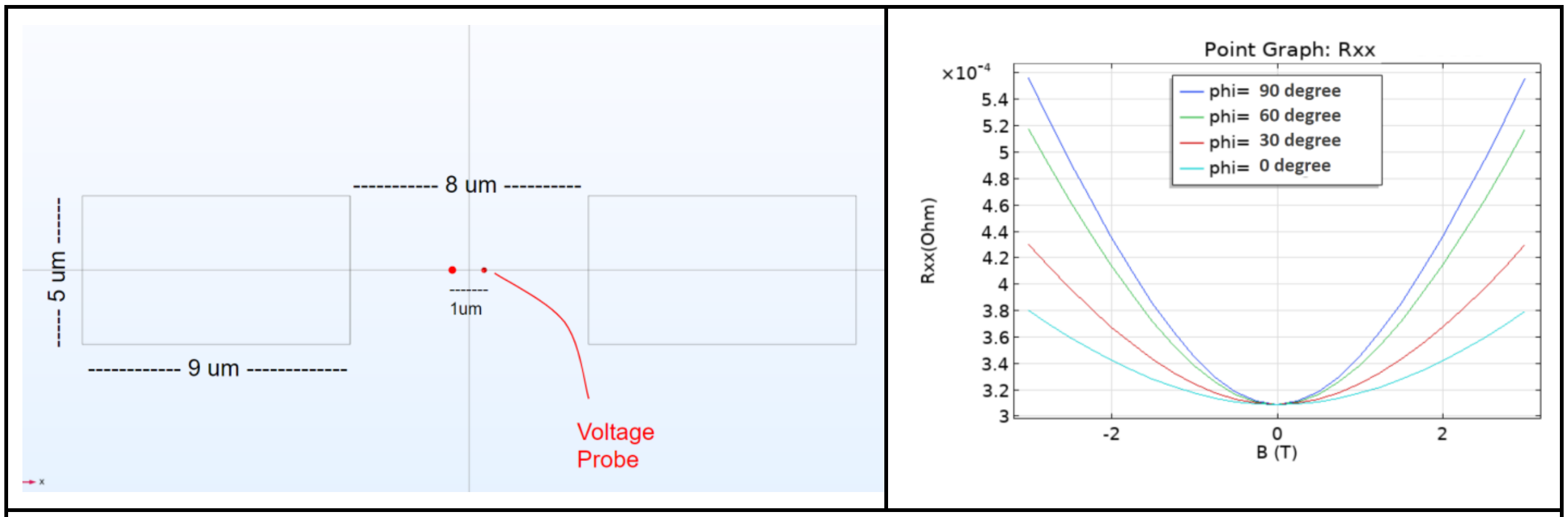

*Figure S2: a) Electrode geometry for 4-probe measurement as seen in the XY plane. b) The longitudinal magnetoresistance as measured at the voltage probes for the 4-probe device geometry shown in a).*

To fully understand the effects of current jetting as a function of the electrode separation, we also calculated current densities for a 4-probe device geometry as shown in Fig. S3a. The voltage probes were kept fixed at a separation of 590 nm, while the current and the ground electrodes were moved further apart on the same face of the metallic cube. Fig. S3b clearly indicates that the current density magnitude in the z-direction drops much faster when the electrodes are closer together. Since the total current is conserved, this indicates that maximal current flows in the desired x-direction when the electrodes are closer together. Fig S4 shows the intensity plots for the current density magnitude as the electrodes are moved apart. On observing the gradient of the intensity in the electrode gap, we further conclude that the current density in the channel is much more uniform when the electrodes are closer together i.e. narrow channel geometry.

An intuitive way to think about this is that separation between the current electrodes set a length scale over which depth the current density is confined primarily to the surface.

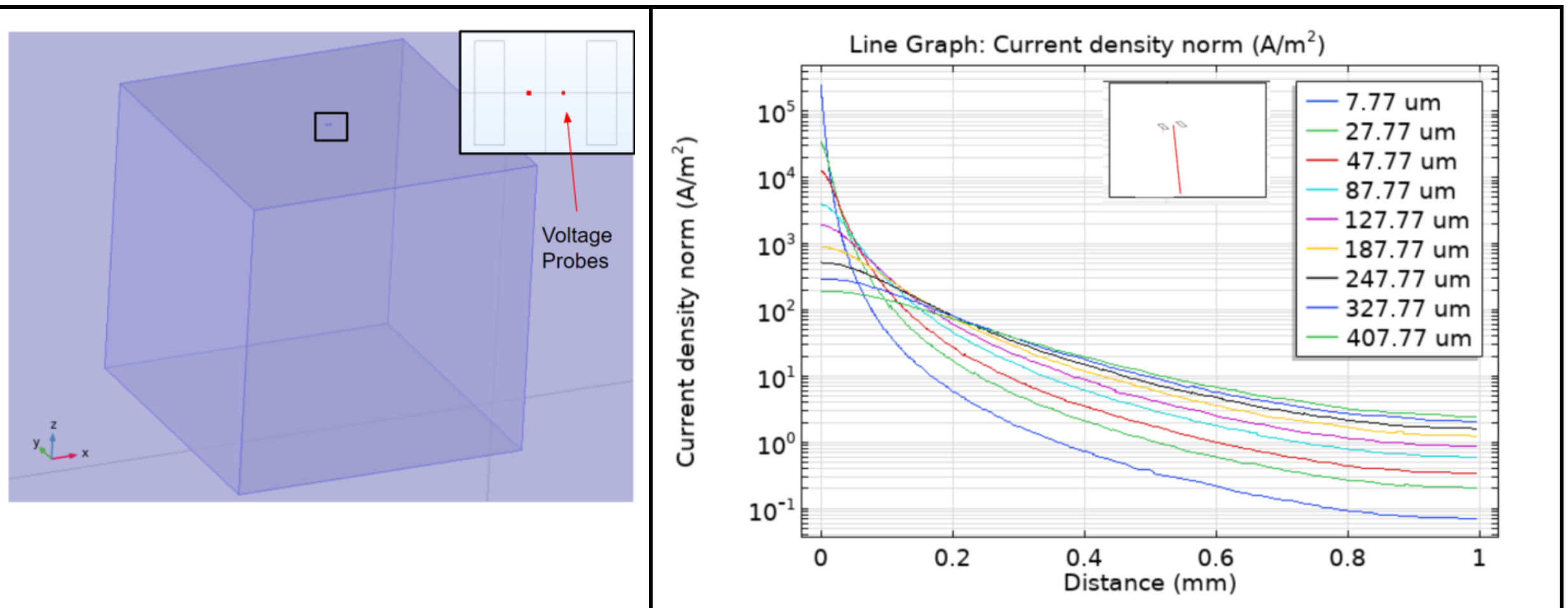

*Figure S3: a) Device geometry for 4-probe measurement as the current and ground electrodes are moved apart. The voltage probes are shown to be points (inset) between which the potential difference is measured. b) Current Density for different electrode separation. The magnitude of the current density as measured along a cut-line in the middle of the voltage probes (inset). The x-axis denotes the distance inside the metal along the cut-line. The magnetic field is kept fixed at 3T, perpendicular to the current direction.*

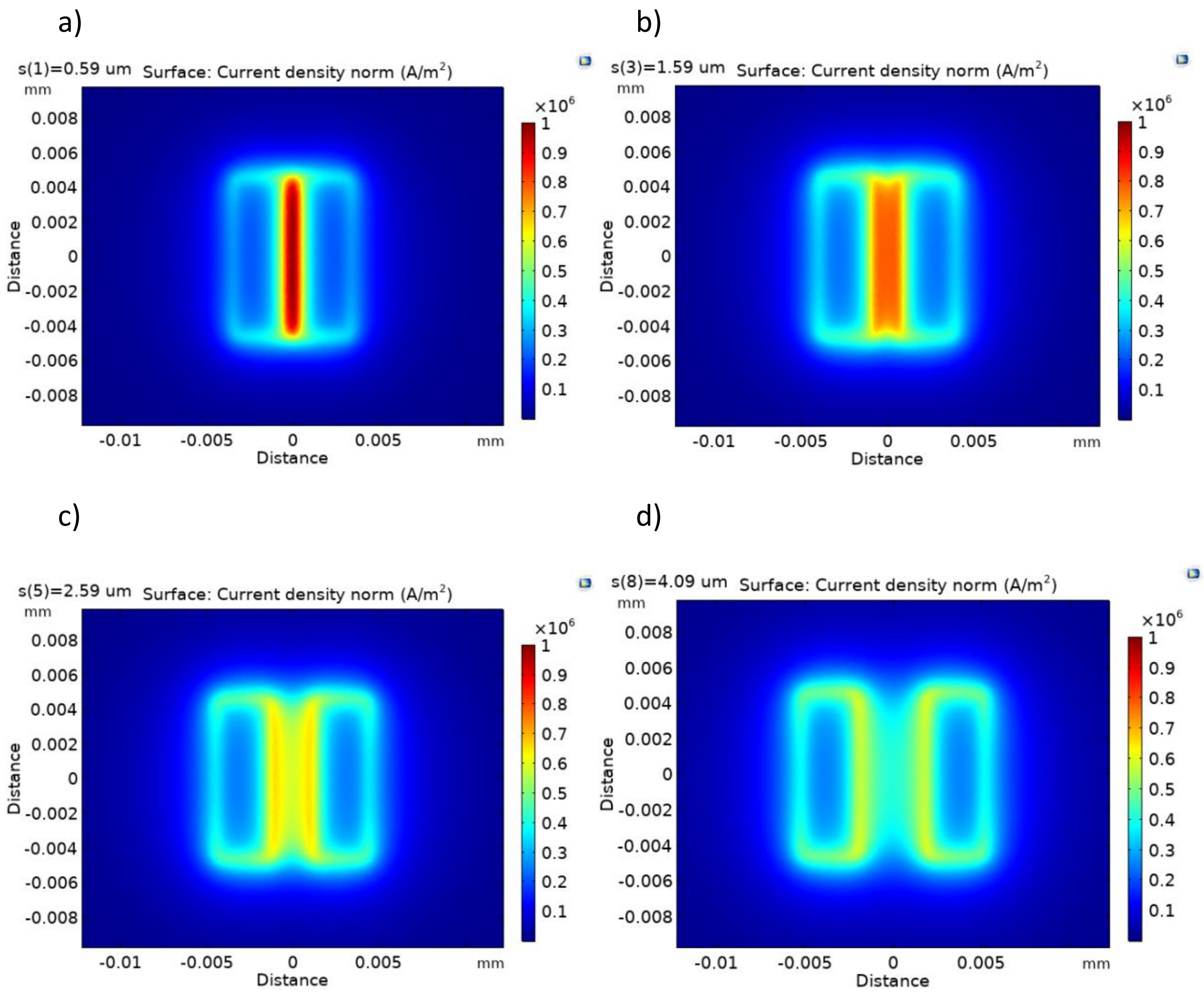

*Figure S4 (a-d): Current density magnitudes in the XY cut-plane at a depth of 1180 nm (twice the separation of the voltage probes) in the z-direction. The magnetic field is kept fixed at 3T, perpendicular to the current direction.*

Finally, we did a simulation to study the current density magnitude for anisotropic magnetoresistance while varying the dimensions of the electrodes. Keeping the injected current and the electrode area fixed, we decreased the width of the electrodes to study the current jetting in the electrode gap. As shown in Fig. S5, if we keep the electrode separation fixed, the current jetting is minimal for wider electrodes. As the electrodes become narrower, we see higher magnitude of current 'spilling' out of the intended path along the x-direction.

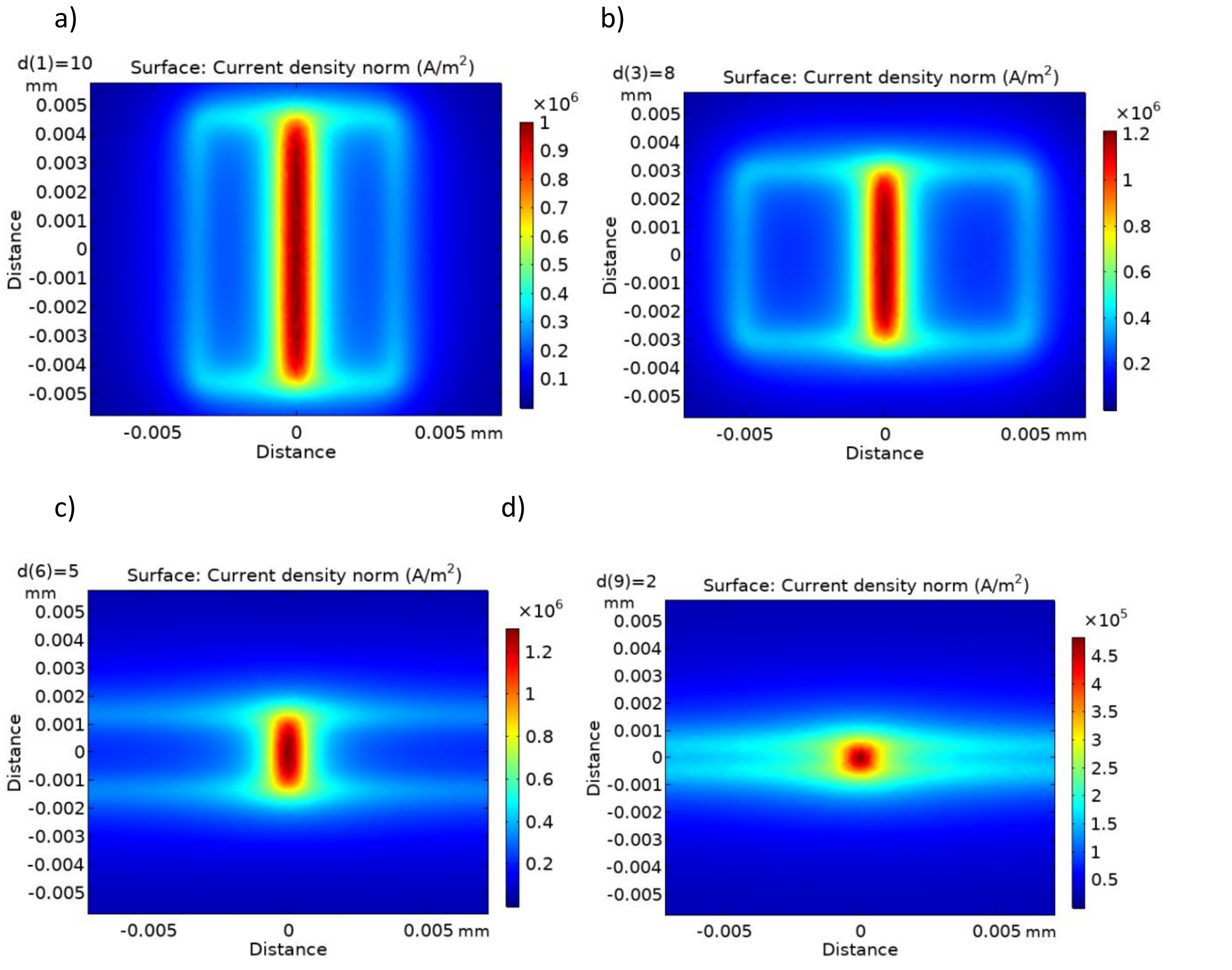

*Figure S5 (a-d): Current density magnitudes in the XY cut-plane at a depth of 590 nm (the separation of the electrodes is kept constant at 590 nm ) in the z-direction. The width and length of the electrodes are varied while keeping the area fixed. The magnetic field is kept fixed at 3T, perpendicular to the current direction.*

We have demonstrated using finite element simulations that our devices with wide current injecting electrodes, nanometer-scale probes, and superconducting contacts are appropriate for a wide variety of studies of quantum crystals, including Weyl semimetals.